\documentclass[12pt]{article}
\usepackage{latexsym}
\usepackage{amssymb}
\usepackage{amsmath}
\usepackage{amsfonts}
\usepackage{amssymb}
\usepackage{amsthm}
\usepackage{indentfirst}
\usepackage{mathrsfs}

\catcode`@=11
\renewcommand{\section}{\@startsection{section}{1}{0pt}{\medskipamount}
{\medskipamount}{\large\bf}} \numberwithin{equation}{section}
\catcode`@=12

\def\m{{\mu}}
\def\n{{\nu}}
\def\a{{\alpha}}

\def\b{{\beta}}
\def\g{{\gamma}}

\def\d{{\delta}}

\def\l{{\lambda}}

\def\vp{{\varphi}}
\def\f{\frac}

\def\la{\label}
\def\eq{\eqref}
\def\pr{\partial}

\newcommand{\be}{\begin{equation}}
\newcommand{\ee}{\end{equation}}

\newcommand{\Tr}{{\rm Tr}}
\newcommand{\sTr}{{\rm sTr}}
\newcommand{\rd}{\overrightarrow{\partial}}
\newcommand{\ld}{\overleftarrow{\partial}}

\def\cD{{\cal D}}
\def\bea{\begin{eqnarray}}
\def\eea{\end{eqnarray}}
\def\nn{\nonumber}

\def\pa{\partial}                       

\def\beq{\begin{eqnarray}}    
\def\eeq{\end{eqnarray}}      

\def\ln{\,\mbox{ln}\,}                  
\def\Tr{\,\mbox{Tr}\,}                  
\def\sTr{\,\mbox{sTr}\,}                
\def\sDet{\,\mbox{sDet}\,}              

\begin{document}

\begin{center}

{\Large\bf Loop expansion of the average effective action in the
functional renormalization  group approach}

\vspace{18mm}

{\Large Peter M. Lavrov$^{(a, b)}\footnote{E-mail:
lavrov@tspu.edu.ru}$,\; Boris S. Merzlikin$^{(a,
c)}\footnote{E-mail: merzlikin@tspu.edu.ru}$\;}

\vspace{8mm}

\noindent  ${{}^{(a)}} ${\em
Tomsk State Pedagogical University,\\
Kievskaya St.\ 60, 634061 Tomsk, Russia}

\noindent  ${{}^{(b)}} ${\em
National Research Tomsk State  University,\\
Lenin Av.\ 36, 634050 Tomsk, Russia}

\noindent  ${{}^{(c)}} ${\em
National Research Tomsk Polytechnical  University,\\
Lenin Av.\ 30, 634050 Tomsk, Russia}

\vspace{20mm}

\begin{abstract}
\noindent We formulate a perturbation expansion for the effective
action in a new approach to the functional renormalization group
method based on the concept of composite fields for regulator
functions being their most essential ingredients. We demonstrate
explicitly the principal difference between the properties of
effective actions in these two approaches existing already on the
one-loop level in a simple gauge model.
\end{abstract}

\end{center}

\vfill

\\


\newpage

\section{Introduction}

The functional renormalization group (FRG) approach
\cite{P,W,W2,Wett-Reu-1,Wett-Reu-2} is a very popular method (see
the recent review in Ref. \cite{G} and references therein) to study
quantum properties of physical models beyond the perturbation
theory. The application of this method to gauge systems meets
essential difficulties which are connected with gauge dependence of
average effective action even on-shell \cite{LS,LS1}. It happens due
to the presence of regulator functions which improve the behavior of
propagators in IR and UV regions but destroy the gauge invariance of
initial classical action. It was the main reason in Refs.
\cite{LS,LS1} to reformulate the standard FRG approach preserving
their attractive features with regulator functions in a way leading
to gauge independence of effective action on its extremals. This is
achieved when regulator functions are considered as composite fields
introducing  on the quantum level with the help of additional
sources. In quantum field theory (QFT), the effective action with
composite fields was introduced and studied within the perturbation
theory by Cornwell, Jackiw and Tomboulis \cite{CJT}. Later, it was
shown that the effective action with composite fields in Yang-Mills
theories \cite{L} as well as in general gauge theories \cite{LO}
does not  depend on gauge on its extremals. This allows one to
consider quantum methods based on the idea of composite fields as
consistent ones. Namely, this fact was the basis for a new approach
to FRG \cite{LS,LS1}.

In the present article, we study the properties of average effective
actions, both in the standard and  new FRG approaches in a loop
approximation. Here, it should be noted that the FRG approach has
been proposed as a method to study nonperturbative quantum effects
with the help of the so-called FRG flow equation for the average
effective action. On the other hand, all renormalization procedures
in QFT are known in the framework of perturbation theory only. In
particular, this means that any approach to the quantum description
of models in QFT might be tested  on the level of perturbation
theory to satisfy some physical requirements. Among such
requirements, the gauge independence of the effective action
on-shell is very essential. Because of this circumstance, in the
present paper, we restrict ourselves to the study of properties of
the average effective action proposed in Refs. \cite{LS,LS1} in the
loop approximation. We find by explicit calculations the difference
existing between the one-loop average effective actions in the
standard and new FRG approaches already in the case of a simple
gauge model. Moreover the average effective action found in this
model is exact in the case of the standard FRG approach without
referring to the perturbation theory and to the flow equation.
\\

\section{Average effective action in the standard FRG approach}

We consider a Yang-Mills theory of fields $A^a_{\mu}$ with the
action $S_0=S_0(A)$ and assume its invariance under the gauge
transformations,
 \beq \label{S0}
S_{0}(A)\;\frac{\overleftarrow{\delta}}{\delta
A^a_{\mu}}\;D^{ab}_{\mu}=0,\quad \delta A^a_{\mu}=D^{ab}_{\mu}\xi^b,
 \eeq
where $D^{ab}_{\mu}=\d^{ab}\pa_{\mu}+f^{acb}A^c_{\mu}$ is the
covariant derivative, $\xi^a$ is an arbitrary gauge function and
$f^{abc}$ are structure constants of a Lie group. Quantization of
the model via the Faddeev-Popov method \cite{FP} involves the
configuration field space
 \beq \label{2.1}
\vp^A=\{A^a_{\mu},B^a,C^a,{\bar C}^a\},
 \eeq
including the ghost ($C^a$) and antighost (${\bar C}^a$) fields and auxiliary fields
($B^a$) with the following distribution of Grassmann parities
 \beq
\label{2.2} \varepsilon(\vp^A)=\varepsilon_A,\quad
\varepsilon(A^a_{\mu})=\varepsilon(B^a)=0, \quad
\varepsilon(C^a)=\varepsilon({\bar C}^a)=1\,.
 \eeq
The Faddeev-Popov action, $S_{FP}(\Phi)$, can be presented in the form
 \beq
\label{2.3}
S_{FP}(\vp)=S_0(A)+\Psi(\vp)\overleftarrow{d},
 \eeq
where the
nilpotent differential $\overleftarrow{d}$
 \beq
 \label{2.4}
 \overleftarrow{d}=\frac{\overleftarrow{\delta}}{\delta
 A^a_{\mu}}D^{ab}_{\mu}C^b+ \frac{\overleftarrow{\delta}}{\delta
 {\bar C}^a}B^a+ \frac{\overleftarrow{\delta}}{\delta
 C^a}\frac{1}{2}f^{abc}C^c C^b, \quad \overleftarrow{d}^2=0,
 \eeq
generates the Becchi-Rouet-Stora-Tyutin (BRST) transformation \cite{brs1,t}
 \beq
\label{2.5}
\delta \vp^A=\vp^A\overleftarrow{d}\mu.
 \eeq
Here, $\Psi(\vp)$ is a gauge fixing Fermion functional,  and $\mu$
is a constant Grassmann parameter. Usually,  the
 \beq
\label{2.6} \Psi(\vp)={\bar C}^a\chi^a(\vp)
 \eeq
form of $\Psi(\varphi)$  is used. One of the more popular choices of
gauge functions, $\chi^a(\vp)=\chi^a(A,B)$, reads
 \beq
\chi^a(A,B)=\pa^{\mu}A^{a}_{\mu}+\frac{\alpha}{2} B^a,
 \eeq
where $\alpha$ is a gauge parameter. The action $S_{FP}(\vp)$ is BRST invariant,
\beq
S_{FP}(\vp)\overleftarrow{d}=0.
\eeq

The main idea of the standard formulation of the FRG approach  is to
modify from the very beginning propagators of vector fields as well
as ghost and antighost fields by introducing the regulator
Lagrangians with a momentum-shell parameter $k$,
 \beq \label{2.7}
 &&L^1_k(x) =\frac{1}{2}
 A^{a\mu}(x)(R_{k,\,A})^{ab}_{\mu\nu}(x)A^{b\nu}(x)\,, \\
 \label{2.8}
 &&L^2_k(x) ={\bar C}^a(x)(R_{k,\,gh})^{ab}(x)C^b(x)=
  {\bar C}^a(x)(\bar R_{k,\,gh})^{ab}(x)C^b(x)\,, \\
&&(\bar R_{k,\,gh})^{ab}(x)= \f12
\Big((R_{k,\,gh})^{ab}(x)-R_{k,\,gh})^{ba}(x)  \Big)\;,
  \eeq
where regulator functions $R_{k,A}$ and $R_{k,gh}$ do not depend
on the fields and obey the properties
 \beq
 \lim_{k\rightarrow 0}(R_{k,\,A})^{ab}_{\mu\nu}\,=\,0
  \,,\qquad
 \lim_{k\rightarrow 0} (R_{k,\,gh})^{ab} \,=\,0\,.
 \eeq
The  generating functional of the Green function is constructed in
the form of a path integral 
\beq
 Z_k(J)= \f{1}{N} \int {\cal D} \varphi\,
 \exp \left\{\frac{i}{\hbar}\,\big[S_{FP}(\varphi)+S_k(\varphi)+J_A\varphi^A)
 \big]\right\}
 \label{2.9}
 =\exp\left\{\frac{i}{\hbar}W_k(J)\right\},
\eeq where $W_k(J)$ is the generating functional of the connected
Green functions, $J_A=J_A(x),\;\varepsilon(J_A)=\varepsilon_A$,
$N$ is a normalization constant,
\beq \label{2.10}
 N = \int {\cal D} \varphi \,
 \exp\left\{\f{i}{2\hbar}\varphi^A\,(iD^{-1}_{AB})\,\varphi^B\right\} =
 \left(\sDet(iD^{-1})\right)^{-\frac{1}{2}},
 \eeq
and
 \beq
\label{2.11}
iD^{-1}_{AB}=\overrightarrow{\pa}_A \big(S_{FP}(\varphi)+
S_k(\varphi)\big)\overleftarrow{\pa}_B\Large|_{\varphi=0},\quad
\pa_A=\frac{\delta}{\delta \varphi^A} .
 \eeq
In Eqs.(\ref{2.9}) and (\ref{2.11}), $S_k(\vp)$ is the regulator
action,
  \beq
  \label{2.12} S_k(\vp)\equiv\frac{1}{2}
\varphi^A\big((L^{1}_k{}'')_{AB}+(L^{2}_k{}'')_{AB}\big)\varphi^B=
\int dx \left[L^1_k(x)+L^2_k(x)\right],
 \eeq
where
 \beq
 (L^{i}_k{}'')_{AB}=\overrightarrow{\pa}_A
L^{i}_k(x)\overleftarrow{\pa}_B,\quad i=1,2,
 \eeq
 are constant supermatrices.


The  effective action, $\Gamma_k(\Phi)$,  is defined as the modified
Legendre transform of $W_k(J)$ \cite{G} with respect to $J_A$,
 \beq
\Gamma_k(\Phi)=W_k(J)-J_A\Phi^A - S_k(\Phi),\quad
\frac{\overrightarrow{\delta}}{\delta J_A}W_k(J)=\Phi^A\,,
 \eeq
so that\footnote{We use the same notation $\pr_A$ meaning the derivative over
 field $\Phi_A$.}
 \beq
\Gamma_k(\Phi)\overleftarrow{\pa}_A=-J_A - S_k(\Phi)\ld_A\,.
 \eeq
The  \ $\Gamma_k(\Phi)$ \ satisfies the functional
integrodifferential equation
 \bea
 \exp \left\{ \frac{i}{\hbar}\,\Gamma_k(\Phi) \right\} &=&
\f{1}{N}
\int D\varphi \,\exp \bigg\{\frac{i}{\hbar}
\Big[S_{FP}(\Phi+\varphi) + S_k(\Phi+\varphi) - S_k(\Phi)
 \nn \\
&& - \big(S_k(\Phi)\ld_A\big)\,\varphi^A
- \big(\Gamma_k(\Phi)\overleftarrow{\pa}_A\big)\,\varphi^A\Big]\bigg\}.
\label{GkEq}
 \eea
It has been shown in Ref. \cite{LS} that the
 effective action (\ref{GkEq}) depends on the gauge even on its
extremals, $\Gamma_k(\Phi)\overleftarrow{\pa}_A=0$.
This fact indicates a serious problem with the physical
interpretation of the results obtained for gauge theories in the
framework of the standard FRG method. And this was the main reason
to reformulate the standard FRG approach in the form being free of
gauge dependence on-shell.

\section{Effective action in the new FRG formulation}

The new FRG approach involves   external scalar sources
$\Sigma_1(x)$ and $\Sigma_2(x)$,
$\varepsilon(\Sigma_1(x))=\varepsilon(\Sigma_2(x))=0$. The
generating functional of the Green functions for Yang-Mills theories
with composite fields is introduced as
 \beq
 {\cal Z}_k(J;\Sigma)= \int {\cal D} \varphi\,
 \exp \left\{\frac{i}{\hbar}\,\big[S_{FP}(\vp)+J_A\vp^A +
 \Sigma_i L^i_k(\vp)\big]\right\}= 
 \label{Zcf}
 \exp\left\{\frac{i}{\hbar}{\cal W}_k(J;\Sigma)\right\},
\eeq where ${\cal W}_k(J;\Sigma)$ is the generating functional of
the Green functions in the presence of composite fields. Here we
introduce the following notation
 \beq
 \Sigma_i L^i_k(\vp)=
\int dx \left[\Sigma_1(x) L^1_k(x) + \Sigma_2(x) L^2_k(x)\right].
 \eeq


Using the explicit structure of the regulator Lagrangians  (\ref{2.7}), (\ref{2.8}) and
the definition (\ref{Zcf}) we deduce the relations
 \beq
\label{relZ}
\frac{\overrightarrow{\delta}}{\delta\Sigma_i}{\cal Z}_k=
\frac{\hbar}{2i}\;
\left(\frac{\overrightarrow{\delta}^2}{\delta J_B\delta J_A}\;{\cal Z}_k\right)\;
(L^{i''}_k)_{AB}(-1)^{\varepsilon_B},
 \eeq
or, in terms of ${\cal W}_k$,
 \beq
\label{relW}
\frac{\overrightarrow{\delta}}{\delta\Sigma_i}{\cal W}_k=\frac{\hbar}{2i}
\left[\left(\frac{\overrightarrow{\delta}^2}{\delta J_B\delta J_A}\;{\cal W}_k\right)
+\frac{i}{\hbar}\left(\frac{\overrightarrow{\delta}}{\delta J_B}{\cal W}_k\right)
\left(\frac{\overrightarrow{\delta}}{\delta J_A}{\cal W}_k\right)\right]
(L^{i''}_k)_{AB}(-1)^{\varepsilon_B}.
 \eeq

The effective action with composite
fields, $\Gamma_k=\Gamma_k(\Phi;F)$, can be introduced by means of
the double Legendre transformations 
 \beq
 \label{Gcf1}
 \Gamma_k(\Phi;F) &=& {\cal W}_k(J;\Sigma)
 \,-\,J_A\Phi^A\,-\Sigma_i\big[L^i_k(\Phi)+ \f12\hbar F^i\big]\,,
 \eeq
where 
 \beq \label{Gcf2}
 \frac{\overrightarrow{\delta}}{\delta J_A}{\cal W}_k(J;\Sigma) = \Phi^A
 \,,\quad  \frac{\overrightarrow{\delta}}{\delta\Sigma_i}{\cal W}_k(J;\Sigma)=
L^i_k (\Phi)+ \f12 \hbar F^i \,,\quad i=1,2.
 \eeq
 From Eq. (\ref{Gcf1}) and (\ref{Gcf2}) it follows that
 \beq
\label{ConGkcf}
 \Gamma_k(\Phi;F) \overleftarrow{\pa}_A=-J_A
 -\Sigma_i\left(L^i_k(\Phi)\overleftarrow{\pa}_A\right) \,,\quad
\Gamma_{k,\,i}(\Phi;F)=-\f12\hbar\,\Sigma_i,\quad
\Gamma_{k,\,i}=\Gamma_k\frac{\overleftarrow{\delta}}{\delta F^i}.
 \eeq

Let us introduce
the full sets of fields ${\cal F}^{\cal A}$ and sources ${\cal
J}_{\cal A}$ according to
 \beq {\cal F}^{\cal A}=(\Phi^A, F^i)
 \,,\qquad {\cal J}_{\cal A}=(J_A,\Sigma_i).
 \eeq

From the condition of the solvability of Eqs. (\ref{ConGkcf}) with
respect to the sources \ $J$ \ and \ $\Sigma$, it follows that
 \beq
 \label{consol}
 \left(\frac{\overrightarrow{\delta }}{\delta {\cal J}_{\cal B}}\,\,
 {\cal F}^{\cal C}({\cal J})\right)
 \left(\frac{\overrightarrow{\delta}}{\delta{\cal F}^{\cal C}}\;{\cal J}_{\cal
 A}({\cal F})\right) \,=\,\delta^{\cal B}_{\cal A}.
 \eeq
One can express \ ${\cal J}_{\cal A}$ \ as a function of
the fields in the form
 \beq
 {\cal J}_{\cal A}
 \,=\,\left(-\left(\Gamma_k{\overleftarrow{\pa}_A}\right)\,+
 \frac{2}{\hbar}\Gamma_{k,\,i}
 \left(L^i_k(\Phi)\overleftarrow{\pa}_A\right),
 -\frac{2}{\hbar}\Gamma_{k,\,i}\right)
 \eeq
and therefore
 \beq
 \frac{\overrightarrow{\delta}}{\delta{\cal F}^{\cal A}}\;
 {\cal J}_{\cal B}({\cal F})
 = -(G^{''}_k)_{{\cal A}{\cal B}}\,,\qquad
 \frac{\overrightarrow{\delta}}{\delta {\cal J}_{\cal A}} \;{\cal F}^{\cal B}({\cal J})=
 -(G^{''-1}_k)^{{\cal A}{\cal B}}\,.
 \eeq
Here
\beq
(G^{''}_k)_{{\cal A}{\cal B}}=\left(\begin{array}{cc}
(\Gamma^{''}_k)_{AB}-
\frac{2}{\hbar}\Gamma_{k,\,i}(L^i_{k}{}'')_{AB} -
\frac{2}{\hbar}\left(\overrightarrow{\pa}_A\Gamma_{k,\,i}\right)
\left(L^i_k(\Phi)\overleftarrow{\pa}_B\right)&
\frac{2}{\hbar}(\overrightarrow{\pa}_A\Gamma_{k,\,j})\\
(\Gamma_{k,\,i}\overleftarrow{\pa}_B)-\frac{2}{\hbar}(\Gamma^{''}_k)_{ij}
\left(L^j_k(\Phi)\overleftarrow{\pa}_B\right)&
\frac{2}{\hbar}(\Gamma^{''}_k)_{ij}\\
\end{array}\right),
\eeq
and
\beq
&&(\Gamma^{''}_k)_{AB}=\overrightarrow{\pa}_A\Gamma_k\overleftarrow{\pa}_B,\quad
(\Gamma^{''}_k)_{ij}=\frac{\overrightarrow{\delta}}{\delta F^i}\Gamma_k
\frac{\overleftarrow{\delta}}{\delta F^j},\\
&&(G^{''}_k)_{{\cal A}{\cal C}}(G^{''-1}_k)^{{\cal C}{\cal B}}=\delta^{\cal B}_{\cal A},\quad
(G^{''-1}_k)^{{\cal A}{\cal C}}(G^{''}_k)_{{\cal C}{\cal B}}=\delta^{\cal A}_{\cal B}.
\eeq

Let us introduce the supermatrix
\beq
{\cal W}^{{\cal A}{\cal B}}_k=\frac{\overrightarrow{\delta }}{\delta {\cal J}_{\cal A}}
{\cal F}^{\cal B}({\cal J}).
\eeq
Then we have
\beq
\label{calWm}
{\cal W}^{{\cal A}{\cal B}}_k=\left(\begin{array}{cc}
{\cal W}^{A B}_k&
\frac{2}{\hbar}\left({\cal W}^{A j}_k-{\cal W}^{A C}_k
\left(\overrightarrow{\pa}_C L^j_k(\Phi)\right)\right)\\
{\cal W}^{i B}_k&
\frac{2}{\hbar}\left({\cal W}^{i j}_k-{\cal W}^{i C}_k
\left(\overrightarrow{\pa}_C L^j_k(\Phi)\right)\right)\\
\end{array}\right),
\eeq
where
\beq
{\cal W}^{AB}_k=\frac{\overrightarrow{\delta}^2{\cal W}_k}{\delta J_A\delta
J_B}\;,\quad
{\cal W}^{A i}_k=\frac{\overrightarrow{\delta}^2{\cal W}_k}{\delta J_A\delta\Sigma_i}\;,\quad
{\cal W}^{i A}_k=\frac{\overrightarrow{\delta}^2{\cal W}_k}{\delta\Sigma_i\delta J_A}\;,\quad
{\cal W}^{ij}_k=\frac{\overrightarrow{\delta}^2{\cal W}_k}{\delta\Sigma_i\delta\Sigma_j}\;.
\eeq
From (\ref{consol}) and (\ref{calWm}) the following relations hold
\beq
\nonumber &&\left[(\Gamma^{''}_k)_{AC}-
\frac{2}{\hbar}\Gamma_{k,\,i}(L^i_{k}{}'')_{AC} -
\frac{2}{\hbar}\left(\overrightarrow{\pa}_A\Gamma_{k,\,i}\right)
\left(L^i_k(\Phi)\overleftarrow{\pa}_C\right)\right]{\cal W}^{CB}_k+\\
\label{consol1}
&&\qquad\qquad\qquad\qquad\qquad\qquad\qquad\qquad\qquad\qquad+\frac{2}{\hbar}
(\overrightarrow{\pa}_A\Gamma_{k,\,j}){\cal W}^{j B}=-\delta_A^{\;\;B},\\
\nonumber
&&\left[(\Gamma^{''}_k)_{AC}-
\frac{2}{\hbar}\Gamma_{k,\,i}(L^i_{k}{}'')_{AC} -
\frac{2}{\hbar}\left(\overrightarrow{\pa}_A\Gamma_{k,\,i}\right)
\left(L^i_k(\Phi)\overleftarrow{\pa}_C\right)\right]\left({\cal W}^{C j}_k-{\cal W}^{C D}_k
\left(\overrightarrow{\pa}_D L^j_k(\Phi)\right)\right)+\\
\label{consol2}
&&\qquad\qquad\qquad\qquad\qquad\qquad+
\f{2}{\hbar}(\overrightarrow{\pa}_A\Gamma_{k,\,i})\left({\cal W}^{i
j}_k-{\cal W}^{i C}_k
\left(\overrightarrow{\pa}_C L^j_k(\Phi)\right)\right)=0,\\
\label{consol3}
&&\left[(\Gamma_{k,\,i}\overleftarrow{\pa}_C)-\frac{2}{\hbar}(\Gamma^{''}_k)_{ij}
\left(L^j_k(\Phi)\overleftarrow{\pa}_C\right)\right]{\cal W}^{C B}_k+
\frac{2}{\hbar}(\Gamma^{''}_k)_{ij}{\cal W}^{j B}_k=0,
\\
\nonumber
&&\frac{2}{\hbar}\left[(\Gamma_{k,\,i}\overleftarrow{\pa}_C)-
\frac{2}{\hbar}(\Gamma^{''}_k)_{ij}
\left(L^j_k(\Phi)\overleftarrow{\pa}_C\right)\right]\left({\cal W}^{C j}_k-{\cal W}^{C D}_k
\left(\overrightarrow{\pa}_D L^j_k(\Phi)\right)\right)+\\
\label{consol4}
&&\qquad\qquad\qquad\qquad+\left(\frac{2}{\hbar}\right)^2
(\Gamma^{''}_k)_{il}\left({\cal W}^{l j}_k-{\cal W}^{l C}_k
\left(\overrightarrow{\pa}_C L^j_k(\Phi)\right)\right)=-\delta_i^{\;j}.
\eeq

In particular, from (\ref{consol1}) and (\ref{consol3})   we deduce
the presentation for ${\cal W}^{AB}$ in terms of the
effective action $\Gamma_k(\Phi;F)$,
 \bea
 \label{calWnp}
{\cal W}^{AB}_k = - \Big( (\Gamma^{''}_k)_{AB}
 - \frac{2}{\hbar}\Gamma_{k,\,i}(L^{i\,''}_{k})_{AB}
 - \big(\rd_A \Gamma_{k,\,i}\big)(\Gamma_k^{-1})^{ij}
\big(\Gamma_{k,\,j}\ld_B \big) \Big)^{-1}.
 \eea
This allows us to present the relation (\ref{relW}) on the level of
the effective action in a closed form
 \beq
\label{relGamm}
-iF^i ={\cal W}^{AB}_k(L^i_k{}'')_{BA}(-1)^{\varepsilon_A}=
\sTr {\cal W}^{AC}_k(L^i_k{}'')_{CB}.
 \eeq
Finally, we discuss the structure of supermatrices
$(L^i_k{}'')_{AB}$ and  the inverse one. According to Eqs. \eq{2.7}
and \eq{2.8}, we have
 \bea
(L^1_k{}'')_{AB} &=& \left(
                     \begin{array}{ccc}
                       (R_{k,\,A})^{ab}_{\m\n} & 0 & 0 \\
                       0 & 0 & 0 \\
                       0 & 0 & 0 \\
                     \end{array}
                   \right)\,, \\
(L^2_k{}'')_{AB} &=& \left(
                     \begin{array}{ccc}
                       0 & 0 & 0 \\
                       0 & 0 & (\bar R_{k,\,gh})^{ba} \\
                       0 & (\bar R_{k,\,gh})^{ab} & 0 \\
                     \end{array}
                   \right)\,.
 \eea
Then
 \bea
 \Sigma_i (L^i_k{}'')_{AB} = \left(
                     \begin{array}{ccc}
                       \Sigma_1(R_{k,\,A})^{ab}_{\m\n} & 0 & 0 \\
                       0 & 0 &  \Sigma_2(\bar R_{k,\,gh})^{ba} \\
                       0 &  \Sigma_2(\bar R_{k,\,gh})^{ab} & 0 \\
                     \end{array}
                   \right)\,.
 \eea
It is useful to introduce the supermatrix
\bea
 (L_k{}^{''-1})^{AB} = \left(
                     \begin{array}{ccc}
                     (R_{k,\,A}^{-1})_{ab}^{\m\n} & 0 & 0 \\
                       0 & 0 &  (\bar R_{k,\,gh}^{-1})_{ba} \\
                       0 & (\bar R_{k,\,gh}^{-1})_{ab} & 0 \\
                     \end{array}
                   \right)\,,
 \eea
 where
 \beq (R_{k,\,A})^{ac}_{\m\alpha}(R_{k,\,A}^{-1})_{cb}^{\alpha\n}=
\delta^a_b\delta^{\nu}_{\mu},\quad (\bar R_{k,\,gh})^{ac}(\bar
R_{k,\,gh}^{-1})_{cb}=\delta^a_b .
 \eeq
We obtain a useful relation
 \beq \label{ortrel} \Sigma_i (L^i_k{}'')_{AC}(L_k{}^{''-1})^{CB} =
\left(
                     \begin{array}{ccc}
                       \Sigma_1\delta^a_b\delta^{\nu}_{\mu}& 0 & 0 \\
                       0 & \Sigma_2\delta^a_b &0 \\
                       0 & 0& \Sigma_2\delta^a_b \\
                     \end{array}
                   \right)\,.
 \eeq

It has been proven in Ref. \cite{LS} that the functional
$\Gamma_k(\Phi;F)$ does not depend on the gauge on its extremals,
 \be
\Gamma_k(\Phi;F)\overleftarrow{\pa}_A=0\,, \qquad
\Gamma_{k,\,i}(\Phi;F)=0.
\la{qem}
 \ee

\section{Loop approximation}

In this section we consider the procedure of loop expansions for
$\Gamma_k(\Phi;F)$, following mainly Ref. \cite{CJT}. Our starting
point is the relation
 \bea
 \nonumber
&&\exp\Big\{\frac{i}{\hbar}\Gamma_k(\Phi;F)\Big\}=
 \exp\Big\{-\f{i}2\, \Sigma_i F^i \Big\} \times \\
 \label{Gammk}
&& \times\int {\cal D} \vp\,
\exp\Big\{\frac{i}{\hbar}\,\left[S_{FP}(\vp)+J_A(\vp^A-\Phi^A)
+\Sigma_i \big(L^i_k(\vp)-L^i_k(\Phi)\big)\right]\Big\},
 \eea
which follows from Eqs. (\ref{Zcf}), (\ref{Gcf1}), and
(\ref{ConGkcf}). Making the background-quantum splitting
 \be
 \vp\rightarrow \vp + \Phi \,,
 \ee
we present Eq. (\ref{Gammk}) in the form
 \bea
 && \exp\Big\{\frac{i}{\hbar}\bar\Gamma_k(\Phi;F)\Big\}=
 \exp\Big\{- \f{i}2\Sigma_i F^i \Big\}\, \int {\cal D}\vp
 \exp\Big\{ \f{i}{\hbar}\left
 [\f{1}{2}\, \vp^A \Big((S''_{FP})_{AB} + \Sigma_i\, (L^i_{k}{}'')_{AB} \Big)\vp^B
\nn \right.\\
&& \qquad\qquad\qquad\qquad\qquad\qquad -\,(\bar\Gamma_k(\Phi;F)\ld_A)\,\vp^A +
S_{int}(\Phi, \vp)  \Big]\Big\} \,, \la{Gamma_bar}
  \eea
where the  notations
 \beq \la{barG2}
&& \bar\Gamma_k(\Phi; F) = \Gamma_k(\Phi; F) - S_{FP}(\Phi)\,,\\
&&S_{int}(\Phi, \vp)= S_{FP}(\Phi+\vp)-S_{FP}(\Phi)
 - (S_{FP}(\Phi)\ld_A) \, \vp^A
 - \f{1}{2}\, \vp^A (S''_{FP})_{AB} \vp^B,  \la{S}\\
 &&i \cD^{-1}_{AB}(\Phi)= \rd_A S_{FP}(\Phi)\ld_B
 \equiv (S''_{FP})_{AB}\,, \la{fp2}
 \eeq
and the relations (\ref{ConGkcf}) are used.

Then we assume the average effective action in the form
 \bea
 {\bar\Gamma}_k(\Phi; F)=\hbar\,\Gamma^{(1)}_k(\Phi;F) + \Gamma_{k\,2}(\Phi;F)\,.
 \la{Gamma}
 \eea
Here $\Gamma^{(1)}_k(\Phi;F)$ is the one-loop effective action for
the set of fields $\Phi_A$ taking into account composite fields
$F^i$. The term $ \Gamma_{k\,2}(\Phi;F)$ includes all the
two-particle-irreducible vacuum graphs in a theory with vertices
determined by $S_{int}(\Phi,\varphi)$ and propagators set equal to
$F^i$. Note that $\Gamma_{k\,2}(\Phi;F)$ by itself is of order
$\hbar^2$ \cite{CJT}.

To calculate the one-loop contribution $\Gamma_k^{(1)}(\Phi,F)$, we
have to omit in the functional integral (\ref{Gamma_bar}) all terms
of order more than $\vp^2$. Then, we have
 \bea
&&\exp\Big\{i\Gamma^{(1)}_k(\Phi;F)\Big\}=
 \exp\Big\{- \f{i}2 \Sigma_i F^i \Big\}\,   \int {\cal D}\vp
 \exp \Big\{ \f{i}{2\hbar}\, \vp^A \Big((S''_{FP})_{AB}
 + \Sigma_i\,(L^{i}_{k}{}'')_{AB} \Big)\vp^B- \nn \\
&&\qquad\qquad\qquad\qquad\qquad\qquad\qquad\qquad\qquad
-i\,(\Gamma^{(1)}_k(\Phi;F)\ld_A)\, \vp^A \Big\} \,.
\la{}
  \eea
The last term in the exponent of the functional integral  reproduces
one-particle-reducible diagrams and should be omitted in calculating
the vertex functions. More systematically, we have the
representation of the exponent
 \bea
 &&\exp \Big\{-i(\Gamma^{(1)}_k(\Phi;F)\ld_A)\,\vp^A \Big\}
 = 1-i\,\left(\Gamma^{(1)}_k(\Phi;F)\ld_A\right)\,\vp^A  \nn \\
 &&\qquad\qquad\qquad\qquad-\f{1}{2}\, \left( \Gamma^{(1)}_k(\Phi;F)\ld_A \right)
 \left(\Gamma^{(1)}_k(\Phi;F)\ld_B\right)
 \vp^B\vp^A + \ldots \;.\la{Gted}
 \eea
After integration over $\varphi^A$, the first term on the right-hand
side \eq{Gted} takes the one-loop contribution to the average
effective action; the second term vanishes  as the Gaussian integral
of an odd function; the third term is responsible for the
cancelation of tadpole diagrams. As a result we arrive at the
relation
 \beq
  \Gamma^{(1)}_k(\Phi;F) - \Gamma^{(1)}_{k,\,i} F^i
 = \f{i}{2} \sTr \ln\Big(i{\cal D}^{-1}_{AB}(\Phi)
 -2 \Gamma^{(1)}_{k,\,i} (L^{i}_{k}{}'')_{AB}  \Big)\,.
 \eeq
At the lower order in $\hbar$, the relation (\ref{relGamm}) reads
 \beq
  \label{Fi} \Big\{ \Big(i {\cal D}^{-1}_{AB}(\Phi)
 - 2\Gamma^{(1)}_{k,\,j}(L^j_k{}'')_{AB}\Big)^{-1} (L^i_k{}'')_{BA} \Big\}
 (-1)^{\varepsilon_A}= - i F^i\,.
 \eeq
From Eq. (\ref{Fi}),  it follows that
 \beq
  \Gamma^{(1)}_{k,\,j}(L^{j}_{k}{}'')_{AB} =
 -\f{i}{2}n_j(F^j)^{-1}(L^{j}_{k}{}'')_{AB}
 + \f{i}{2} {\cal D}^{-1}_{AB}(\Phi)\,, \la{S1}
\eeq where \footnote{Here we don not discuss a suitable definition
of the functional traces, but we assume their existence only.}
 \beq
n_1=\Tr\delta^{\mu}_{\nu}\delta^b_a,\quad n_2=-2\Tr \delta^b_a .
 \eeq
Then, we find
  \beq
 &&\Gamma^{(1)}_{k,\,1}=-\frac{i}{2}n_1 (F^1)^{-1}+\frac{1}{2n_1}
 \Tr (i{\cal D}^{-1})^{ac}_{\mu\alpha}
 (R^{-1}_{k,A})^{\alpha\nu}_{cb},\\
 &&\Gamma^{(1)}_{k,\,2}=\frac{i}{4}n_2 (F^2)^{-1}
 +\frac{1}{2n_2}\Tr (i{\cal D}^{-1})^{ac}
 (\bar R^{-1}_{k,gh})_{cb}.
 \eeq
These relations can be presented in the  form
 \beq
 \label{newpr} \Gamma^{(1)}_{k,\,j}=-\frac{i}{2}m_j
 (F^j)^{-1}+\frac{1}{2m_j}\sTr i{\cal D}^{-1}_{AC}(L_k{}^{''-1})^{CB},
 \eeq
where in the first term on the left-hand side of (\ref{newpr}) there
is no summation over index $j$ and $n_1=m_1, n_2=-2m_2$,
 \beq
 \Gamma^{(1)}_k(\Phi;F)
 = \f{1}{2 m_j} \sTr i{\cal D}^{-1}_{AC}(L_k{}^{''-1})^{CB}\, F^j
  + \f{i}{2} \sTr\ln\big(i n_j(F^j)^{-1}(L^{j}_{k}{}'')_{AB}\big)
  + {\rm const} \la{Gamma1}\,,
 \eeq
where "const" is used to collect all terms independent on the background fields.

Let us consider the equation for  $\Gamma_{k\,2}(\Phi;F)$,
 \beq
 \nonumber
 &&\Gamma_{k2}(\Phi;F)-\Gamma_{k2\;,j}(\Phi;F)F^j=
 -\hbar\left(\Gamma^{(1)}_k(\Phi;F)-\Gamma^{(1)}_{k\,,j}(\Phi;F)F^j\right) -\\
 &&-i\hbar \ln \int {\cal D} \vp
 \exp\bigg[\f{i}{2\hbar} \,
 \vp^A \bigg(i{\cal D}^{-1}_{AB}(\Phi) +
 \big(-2\Gamma^{(1)}_{k,j}(\Phi;F)-
 \frac{2}{\hbar}\Gamma_{k2\;,j}(\Phi;F)\big) (L^{j}_k{}'')_{AB} \bigg)\vp^B \nn\\
 &&\qquad\qquad\qquad
 - \frac{i}{\hbar}\, \left({\bar\Gamma}_{k}(\Phi;F)\ld_A\right)\,\vp^A  +
 \f{i}{\hbar}\,S_{int}(\Phi, \vp)\bigg]\,,
 \eeq
or, taking into account Eqs. (\ref{S1}), (\ref{newpr}), (\ref{Gamma1}), one can
rewrite the last equation in the form
 \beq
\nonumber &&\Gamma_{k2}(\Phi;F)-\Gamma_{k2\;,j}(\Phi;F)F^j =
-\frac{i\hbar}{2}
\sTr\ln\left[in_j(F^j)^{-1}(L^{j}_k{}'')_{AB}\right] \\
&&\qquad\qquad
 -i\hbar \ln \int {\cal D} \vp \exp\bigg[\f{i}{2\hbar}\,
 \vp^A \bigg(in_j(F^j)^{-1}
-\frac{2}{\hbar}\Gamma_{k2\;,j}(\Phi;F)\bigg) (L^{j}_k{}'')_{AB}\vp^B \nn\\
\label{Gamk2}
&&\qquad\qquad\qquad
- \frac{i}{\hbar}\, \left({\bar\Gamma}_{k}(\Phi;F)\ld_A\right)\,\vp^A  +
\f{i}{\hbar}\,S_{int}(\Phi, \vp)\bigg]\,,
 \eeq
Further analysis of this equation requires the explicit form of
$S_{int}(\Phi,\vp)$ supported by the additional restriction on
$\Gamma_{k2\;,j}(\Phi;F)$, which comes from the consistency
condition (\ref{relGamm}).  We are going to study in the future
these equations and their solutions using some special field models.

\section{Gauge (in)dependence: A simple  example}

In this section we illustrate the problem of gauge dependence using
a simple example. To this end, we consider the average effective
action ${\bar\Gamma}_k(\Phi; F)$ up to first order in $\hbar$,
 \beq
\label{Gk1} {\bar\Gamma}_k(\Phi; F)=\hbar\Gamma^{(1)}_k(\Phi; F),
 \eeq
where $\Gamma^{(1)}_k(\Phi; F)$ is defined in Eq. (\ref{Gamma1}).
Note that in consistent gauge theories the effective action does not
depend on  the gauge on its extremals. First, we check
 the gauge dependence of the effective action (\ref{Gk1}). Consider
 the quantum equations of motion $\Gamma^{(1)}_{k,j}(\Phi; F)=0$.
Because of Eqs. \eq{S1} and \eq{newpr}, we have
 \bea
 && -\f{i}{2}n_j(F^j)^{-1}(L^{j}_{k}{}'')_{AB}
 + \f{1}{2} i{\cal D}^{-1}_{AB}(\Phi) = 0\,, \nn \\
 && -\frac{i}{2}m_j (F^j)^{-1}+\frac{1}{2m_j}\sTr i{\cal
 D}^{-1}_{AC}(L_k{}^{''-1})^{CB} = 0\,.
 \la{S=0}
 \eea
 Substituting \eq{S=0} into
\eq{Gamma1} and keeping in mind the definition \eq{fp2}, we obtain
 \bea
 \label{Gkem}
 \Gamma^{(1)}_k(\Phi;F) = \f{i}{2} \sTr\ln S''_{FP}(\Phi)\,.
 \eea
In this approximation, the average effective action (\ref{Gkem})
coincides with the one-loop answer for effective action in a given
Yang-Mills theory. It is well-known fact (see, for example, Ref.
\cite{BarV}) that it does not depend on the gauge when the fields
$\Phi^A$ satisfy the quantum equations of motion. The one-loop
contribution to the average effective action,
$\Gamma^{(1)}_k(\Phi)$, in the standard FRG  approach reads
 \bea
\Gamma^{(1)}_k(\Phi) = \f{i}{2}\sTr\ln\big(S''_{FP}(\Phi) + S''_k(\Phi) \big)\,.
\la{one-loop FRG}
 \eea
This action depends on the gauge even on its extremals. To
illustrate this feature explicitly, we restrict ourselves to the
case of the electromagnetic field in  flat space-time.
The classical action of the model is
 \bea
 \label{mod1}
 S_0(A)=-\f14 \int d^4 x\, F_{\m\n} F^{\m\n}\,, \qquad F_{\m\n} = \pr_\m
 A_\n - \pr_\n A_\m\,.
 \eea
We choose the gauge fixing function in the form
 \bea
 \label{mod2}
 \chi(A,B) = \f{1}{\sqrt{1+\l}}\, \pr^\a A_\a + B\,. \la{chi}
 \eea
Integrating over field $B$ yields the gauge fixing action
 \bea
 \label{mod3}
 S_{gf}(A) = - \f{1}{2(1+\l)} \int d^4 x\, (\pr^\a A_\a)^2\,.
 \eea
The action for ghosts reads
 \bea
 \label{mod4}
 S_{gh}(\bar C, C) = \f{1}{\sqrt{1+\l}} \int d^4 x\, \bar C (\pr^\a \pr_\a) C\,.
 \eea
The effective action of the model in the standard approach \cite{FP}
to gauge theories is \footnote{In this case $\Phi^A=(A, \bar C,
C)$.}
 \beq
 \Gamma(\Phi) = S(\Phi) + i\hbar\, \Gamma^{(1)}(\l),\quad
S(\Phi)=S_0(A)+S_{gf}(A) + S_{gh}(\bar C, C), \la{ea}
 \eeq
 where
 \beq
 \Gamma^{(1)}(\l)&=& \f{1}{2}\Tr \ln\bigg(\square \d^\a_\b
 -\f{\l}{1+\l}\pr^\a\pr_\b \bigg)
 - \Tr\ln\bigg(\f{1}{\sqrt{1+\l}}\,\square\bigg)\,.
 \la{1lea}
 \eeq

The dependence of the effective action $\Gamma(\Phi)$ (\ref{ea}) on
the gauge parameter $\lambda$ is described by the relation
 \beq
\delta\Gamma(\Phi)= \f{\d S(\Phi)}{\d \Phi } \d \Phi +
i\hbar\frac{\pa \Gamma^{(1)}(\lambda)}{\pa\lambda}\delta\lambda.
 \eeq
Using the quantum equations of motion, which in our case coincide
with classical ones
 \bea
 \f{\d \Gamma(\Phi)}{\d \Phi} =  \f{\d S(\Phi)}{\d \Phi} =0\,,
 \eea
we see that all dependence on $\lambda$ comes from
$\Gamma^{(1)}(\lambda)$. In turn,
 \bea
\Gamma^{(1)}(\l)&=& \Gamma^{(1)}(0)
 +\f{1}{2}\Tr \ln\bigg(\d^\a_\b
 -\f{\l}{1+\l}\f{\pr^\a\pr_\b}{\square} \bigg)
 - \ln\f{1}{\sqrt{1+\l}}\,\Tr{\bf 1} \nn \\
&=&\Gamma^{(1)}(0)
 +\f{1}{2}\ln\f1{1+\l} \Tr \f{\pr^\a\pr_\b}{\square}
 -\ln\f{1}{\sqrt{1+\l}}\,\Tr{\bf 1}
= \Gamma^{(1)}(0) \,,
 \eea
where the relation
$
 \Tr \big(\f{\pr^\a\pr_\b}{\square}\big) = \Tr{\bf 1}
 $
is used. Therefore,
 \beq \delta\Gamma(\Phi)\Big|_{ \f{\d
\Gamma(\Phi)}{\d \Phi}\,= 0}=0.
 \eeq
According to Eq. (\ref{Gkem}), the same result is valid for the
average effective action in the new FRG approach \cite{LS,LS1}.

Calculation of the one-loop  effective action of the model within
the standard FRG method gives
 \beq
 \Gamma_k(\Phi) = S(\Phi) + i\hbar\,\Gamma^{(1)}_k(\l),    \la{ea2}
 \eeq
where the action $S(\Phi)$ is defined in Eq. (\ref{ea}). The
regulator action $S_k(A,\bar C, C)$  for the model under
consideration has the form
 \beq
S_k(\Phi)= \f12 \int d^4x\, A^\a (R_{k,\,A})_{\a\b}A^\b
+ \int d^4 x\, \bar C R_{k,\,gh} C,
 \eeq
and the one-loop contribution \eq{one-loop FRG},
$\Gamma^{(1)}_k(\l)$, reads
 \beq
\Gamma^{(1)}_k(\l) =
 \f{1}{2}\Tr \ln\bigg(\square \d^\a_\b
 -\f{\l}{1+\l}\pr^\a\pr_\b + (R_{k,\,A})^\a_\b \bigg)
- \Tr\ln\bigg(\f{1}{\sqrt{1+\l}}\,\square +  R_{k,\,gh}\bigg)
\la{1lea2} \,.
 \eeq

As in the previous case the quantum equations of motion,
 \bea
  \f{\d \Gamma_k(\Phi)}{\d \Phi} =
  \f{\d S(\Phi)}{\d \Phi} = 0\,, \la{eom2}
 \eea
coincide with the classical ones, and the gauge dependence of the
effective action $ \Gamma_k(A)$ (\ref{ea2}) on its extremals comes
essentially from $\Gamma^{(1)}_k(\l)$, which can be presented in the
form
 \bea
 \label{Gkst}
 \Gamma^{(1)}_k(\l) &=& \Gamma^{(1)}(\l)
 + \f{1}{2}\Tr\ln\big(1-G^\a_\g(\l)\,(R_{k,\,A})^\g_\b \big)
 - \Tr\ln\bigg(1 + \sqrt{1+\l}\,\f{R_{k,\,gh}}{\square}\bigg).
 \eea
Here $\eta_{\a\b} = {\rm diag(1,-1,-1,-1)}$ is the Minkowski metric,
$\Gamma^{(1)}(\l) = \Gamma^{(1)}(0)$ is defined in Eq.\eq{1lea}, and
$G^\a_\g(\l)$ is the Green function
 \bea
\left( \square \d^\a_\g - \f{\l}{1+\l}\pr^\a\pr_\g \right)
 G^\g_\b(\l) = - \d^\a_\b \,,\quad
G^\g_\b(\l) = -\f{\d^\g_\b}{\square}- \l\f{\pr^\g
\pr_\b}{\square^2}\,. \la{1lea3}
 \eea
The last two terms on the right-hand side Eq.\eq{Gkst} explicitly
depend on the gauge-fixing parameter $\lambda$. Using the following
property of cutoff functions $R_k(p)\rightarrow 0$ when $k \to 0$,
we can approximate the trace of the logarithm by a linear term:
 \bea
 \Gamma^{(1)}_k(\l) \approx \Gamma^{(1)}(0)
 + \f{1}{2}\Tr\bigg(\f{(R_{k,\,A})^\a_\b}{\square}
 + \l\f{\pr^\a \pr_\g (R_{k,\,A})^\g_\b}{\square^2}
 \bigg) 
- \sqrt{1+\l}\,\Tr\bigg(\f{R_{k,\,gh}}{\square}\bigg)\,.
 \eea
It is clear that
 \beq \frac{\pa \Gamma^{(1)}_k(\l)}{\pa \lambda}\neq 0,
  \eeq
and one meets the gauge dependence of the average effective action
within the standard FRG approach even on-shell.
\\

\section{Discussions}
In this paper we have studied the procedure of loop expansion in the
new FRG approach based on the idea to consider regulator functions
being main ingredients of standard FRG method as composite fields
\cite{LS,LS1}. We have derived an explicit formula at leading order
in $\hbar$ for the average effective action. We have explicitly
demonstrated the gauge dependence of the average effective actions
constructed within the standard and new FRG methods using a simple
gauge model of Abelian vector fields. This example confirmed the
general statement of Refs.\cite{LS,LS1} concerning the gauge
dependence of the standard average effective action even on-shell.
It is very important to note that, in fact, the average effective
action for the model (\ref{mod1}) - (\ref{mod4}) is exact in the
case of the standard FRG approach without referring to perturbation
theory and to solutions of the flow equation. In our opinion this
result indicates at least that the gauge dependence problem within
the standard FRG approach remains open up to now. Perhaps not all
the hidden features of the modified Slavnov-Taylor identities (among
recent studies, see, for example, Ref.\cite{safari}) and the FRG
flow equation are used to respect the BRST symmetry.

The main feature of the FRG approach is its nonperturbative
character, encoding into the FRG flow equation for the average
effective action. This equation is a very complicated nonlinear
functional differential equation for which exact solutions are not
known and  different approximations have been developed (for detail,
see Ref. \cite{G}). In turn the structure of the FRG flow equation
in the new approach \cite{LS,LS1} is also very complicated but
differs from the standard one. This means that solutions to the new
FRG equation require serious efforts to develop new approximation
methods. We plan in the future to present our study of the problem.

\section*{Acknowledgments}
\noindent

The authors  would like  to thank I. L. Shapiro for interesting discussions.
P.M. Lavrov is grateful to the Mainz Institute for Theoretical Physics
(MITP) for its hospitality and its partial support during the completion of this
work. The authors are thankful to the grant of Russian Ministry of
Education and Science, project 2014/387/122 for support. The work is
also supported in part by the Presidential grant 88.2014.2 for
LRSS and DFG grant LE 838/12-2.



\end{document}